\documentclass[fleqn,usenatbib]{mnras}

\usepackage{amsmath}
\usepackage{amssymb}
\usepackage{graphicx}		
\usepackage{comment}

\title[AGN in dwarf galaxies]{AGN in dwarf galaxies: frequency, triggering processes and the plausibility of AGN feedback}

\author[Sugata Kaviraj et al.]{
Sugata Kaviraj,$^{1}$\thanks{E-mail: s.kaviraj@herts.ac.uk}
Garreth Martin$^{1,2}$ and Joseph Silk$^{2,3,4,5}$
\\
$^{1}$Centre for Astrophysics Research, University of Hertfordshire, College Lane, Hatfield AL10 9AB, UK\\
$^{2}$Department of Physics, University of Oxford, Keble Road, Oxford OX1 3RH, UK\\
$^{3}$Institut d'Astrophysique de Paris (UMR 7095: CNRS and UPMC-Sorbonne Universites), 98 bis bd Arago, F-75014 Paris, France\\
$^{4}$Laboratoire AIM-Paris-Saclay, CEA/DSM/IRFU, CNRS, University Paris VII, F-91191 Gif-sur-Yvette, France\\
$^{5}$Department of Physics and Astronomy, The Johns Hopkins University Homewood Campus, Baltimore, MD 21218, USA\\
\\
}

\begin{document}
\label{firstpage}
\pagerange{\pageref{firstpage}--\pageref{lastpage}}
\maketitle

\begin{abstract}
While AGN are considered to be key drivers of the evolution of massive galaxies, their potentially significant role in the dwarf-galaxy regime (M$_{*}<$ 10$^9$ M$_{\odot}$) remains largely unexplored. We combine optical and infrared data, from the Hyper Suprime-Cam (HSC) and the Wide-field Infrared Explorer (WISE) respectively, to explore the properties of $\sim$800 AGN in dwarfs at low redshift ($z<0.3$). Infrared-selected AGN fractions are $\sim$10 -- 30 per cent in dwarfs, which, for reasonable duty cycles, indicates a high BH-occupation fraction. Visual inspection of the deep HSC images indicates that the merger fraction in dwarf AGN ($\sim$6 per cent) shows no excess compared to a control sample of non-AGN, suggesting that the AGN-triggering processes are secular in nature. Energetic arguments indicate that, in both dwarfs and massive galaxies, bolometric AGN luminosities ($L_{AGN}$) are significantly greater than supernova luminosities ($L_{SN}$). $L_{AGN}/L_{SN}$ is, in fact, higher in dwarfs, with predictions from simulations suggesting that this ratio only increases with redshift. Together with the potentially high BH-occupation fraction, this suggests that, if AGN feedback is an important driver of massive-galaxy evolution, the same is likely to be true in the dwarf regime, contrary to our classical thinking.  
\end{abstract}


\section{Introduction}
The evolution of galaxies and their central black holes (BHs) are considered to be intimately linked in our structure-formation model. The apparently generic presence of super-massive black holes (SMBHs) in massive (M$_{*}$$>$10$^9$ M$_{\odot}$) galaxies \citep[e.g.][]{Merritt2001} has led to a paradigm in which (negative) AGN feedback is routinely employed in models to bring key properties of these objects, such as stellar masses and colours, in agreement with observational data \citep[e.g.][]{Beckmann2017}. While the empirical details of how AGN and their host galaxies co-evolve, and how this relationship changes with redshift, remain unclear \citep[e.g.][]{Sarzi2016}, consensus favours a scenario where AGN play an important role in regulating the baryonic content of massive galaxies across cosmic time. 

While past efforts have focused on AGN in massive systems, there is no a priori reason to assume that the presence of central BHs does not continue into the dwarf regime (M$_{*}<$10$^9$ M$_{\odot}$). If the BH-host mass ratio ($\sim$ 1/500) remains similar in dwarfs, these `intermediate-mass' BHs (IMBHs) should have masses less than $\sim$ 10$^6$ M$_{\odot}$. Nevertheless, the potentially important role of AGN in dwarfs has remained largely unexplored \citep{Mezcua2017}, mostly due to a lack of firm evidence that IMBHs are indeed widespread in low-mass systems. 

\begin{figure*}
\center
\includegraphics[width=\columnwidth]{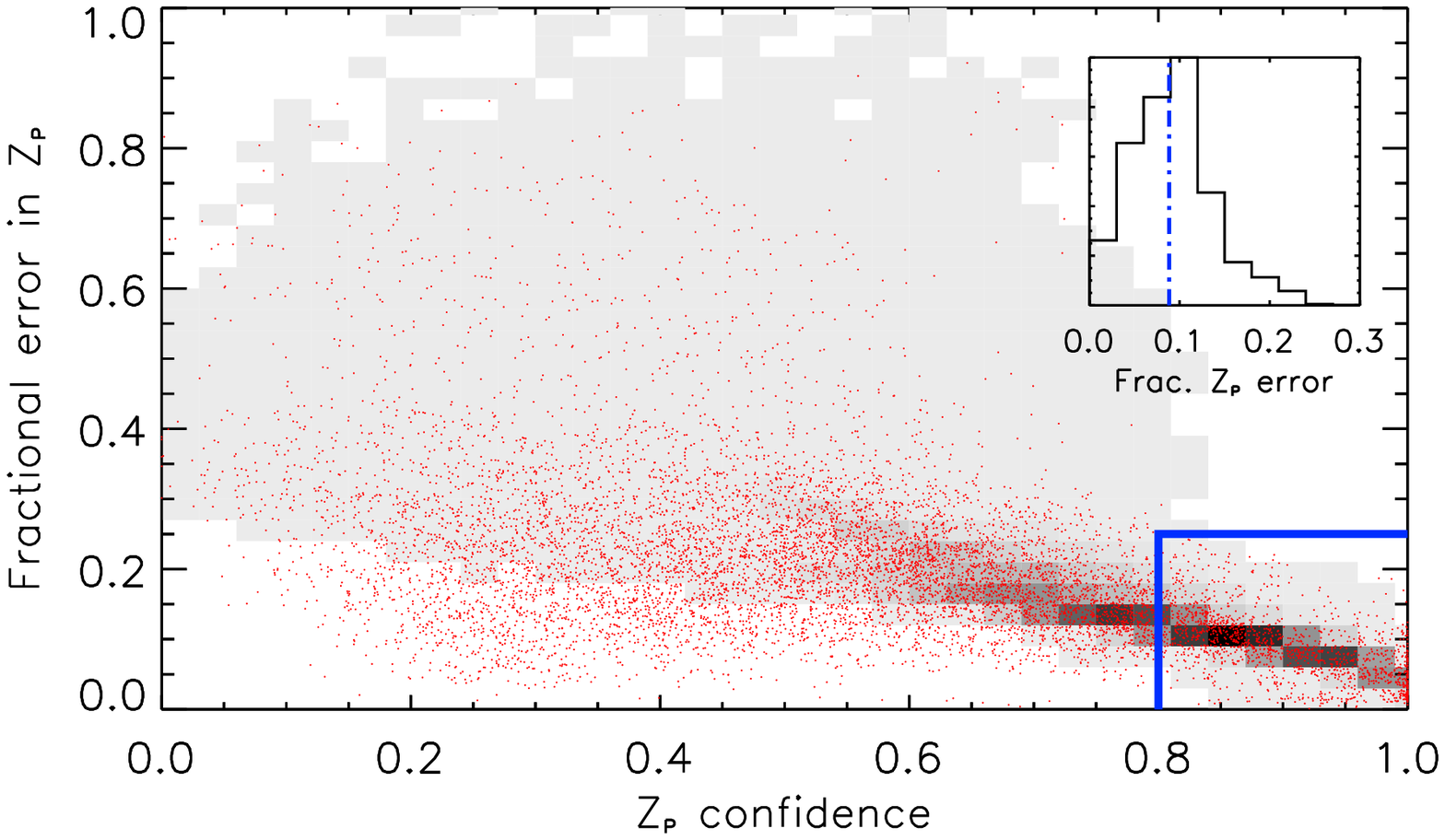}
\includegraphics[width=\columnwidth]{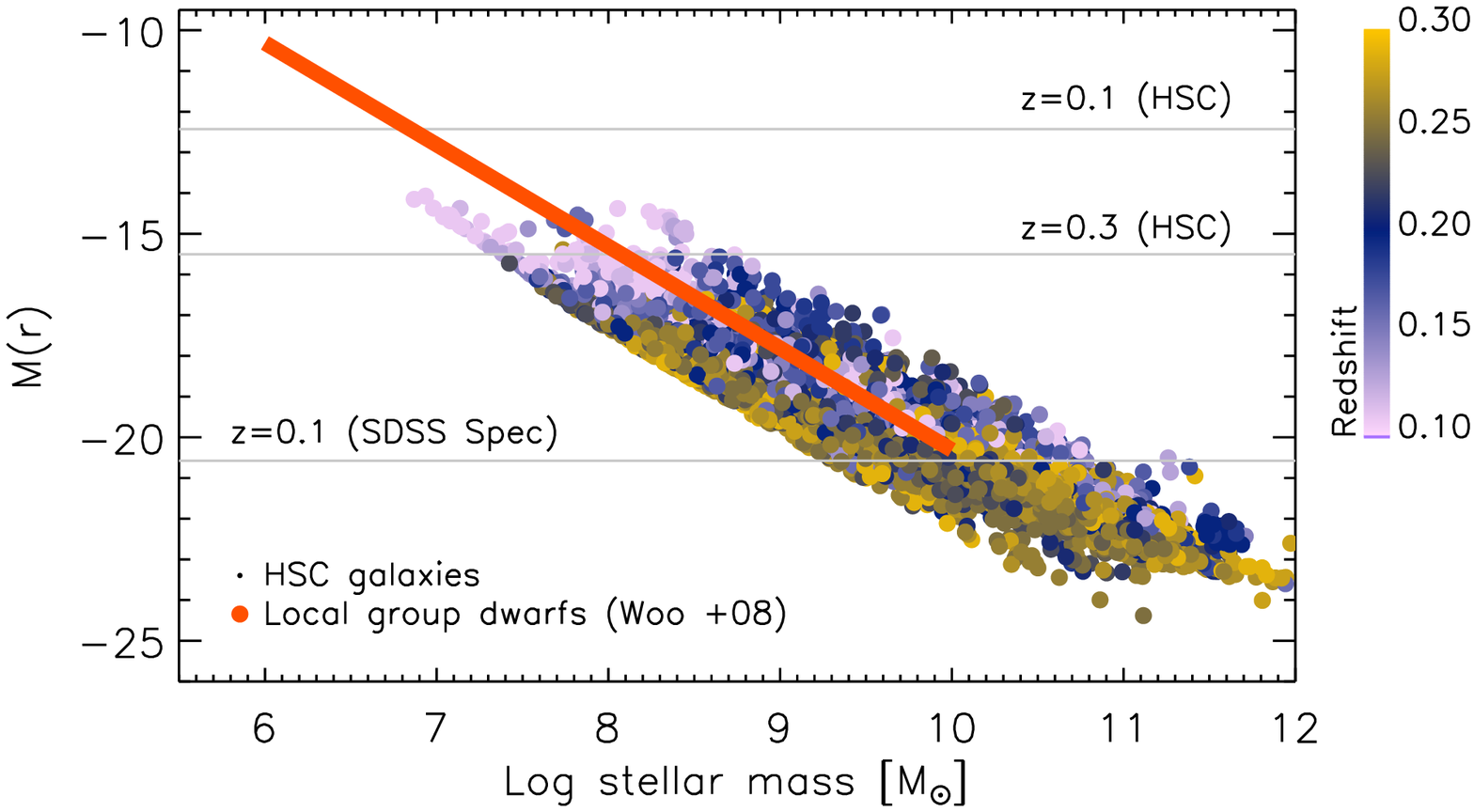}
\caption{\textbf{Left:} The fractional error in HSC photometric redshifts vs the redshift confidence parameter from the HSC catalog. {\color{black}The 2D histogram indicates all galaxies, while the red dots indicate dwarfs (M$_{*}<$10$^9$ M$_{\odot}$)}. The blue rectangle shows the region that is used for this study. The inset shows the distribution of the fractional errors in the photometric redshifts for dwarfs in this region (the median value is $\sim$9 per cent). \textbf{Right:} Absolute $r$-band magnitude vs stellar mass for the HSC-SSP galaxies. Grey horizontal lines indicate completeness limits at various redshifts for the HSC-SSP Wide and SDSS spectroscopic datasets, which have $r$-band detection limits of 26 and 22 mag respectively. Galaxy redshifts are shown colour-coded. The red line shows the median locus of local dwarfs from \citet{Woo2008}, which agrees well with the locus for the HSC galaxies.}
\label{fig:completeness}
\end{figure*}

If dwarfs do generically host IMBHs, as suggested by recent studies \citep[e.g.][]{Reines2013,Mezcua2016,Marleau2017}, then AGN feedback could offer solutions to several outstanding problems in this mass regime \citep[e.g.][]{Silk2017}. For example, the excess number of dwarfs commonly seen in simulations \citep[e.g.][]{Kaviraj2017} cannot be removed via supernova (SN) feedback alone \citep[e.g.][]{Bland-Hawthorn2015}, but could be suppressed by additionally invoking AGN feedback \citep[e.g.][]{Keller2016}. IMBH feedback could help mitigate the `too-big-to-fail' problem, whereby simulated dwarfs tend to be more massive than their observed counterparts \citep[e.g.][]{Garrison-Kimmel2013}. It could also offer an explanation for the significant fraction ($\sim$30 per cent) of baryons that are missing from today's massive disks \citep{Shull2012}, by ejecting material from their dwarf progenitors in the early Universe \citep[e.g.][]{Peirani2012}. IMBH-driven AGN could contribute to re-ionization \citep[e.g][]{Volonteri2009} and provide the seeds for SMBH formation \citep{Johnson2013}, removing the need to invoke super-Eddington accretion to drive SMBH growth at high redshift. 

In this observational study, we use deep optical and infrared data to (a) explore the infrared-selected AGN fraction in nearby dwarfs (\S \ref{sec:agn properties}), (b) study whether the triggering of dwarf AGN is likely to be driven by mergers or secular processes (\S \ref{sec:agn properties}) and (c) compare AGN and SN energetics, to explore the plausibility of AGN feedback in the dwarf regime (\S \ref{sec:ratios}). We summarise our findings in \S \ref{sec:summary}.


\section{Data}
\label{sec:data}

\subsection{The Hyper Suprime Cam (HSC) Subaru Strategic Program and the AllWISE Infrared Catalog}

The HSC Subaru Strategic Program (HSC-SSP) is an imaging survey in $grizy$ and 4 narrow-band filters \citep{Aihara2018a}. HSC offers a 1.5 degree field of view and a median $i$-band seeing of $\sim$0.6 arcseconds, with the HSC-SSP providing three nested layers at varying $r$-band depths (shown in brackets): Wide (26 mag arcsec$^{-2}$), Deep (27 mag arcsec$^{-2}$) and Ultra-deep (28 mag arcsec$^{-2}$). Here, we construct our galaxy sample using the 100 deg$^2$ Wide layer of the HSC-SSP Data Release 1 \citep[DR1;][]{Aihara2018b}. 

\begin{figure*}
\center
\includegraphics[width=\columnwidth]{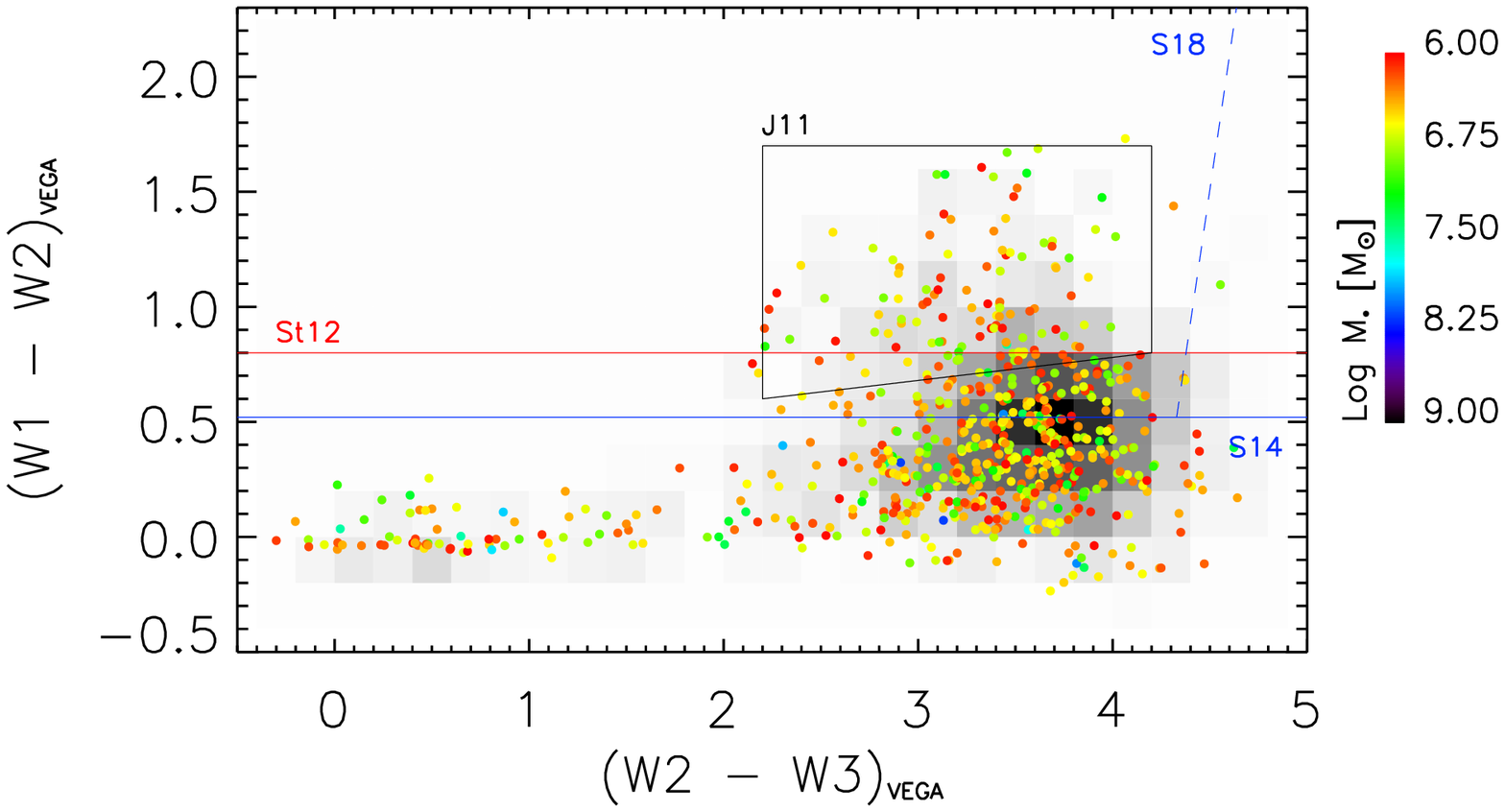}
\includegraphics[width=\columnwidth]{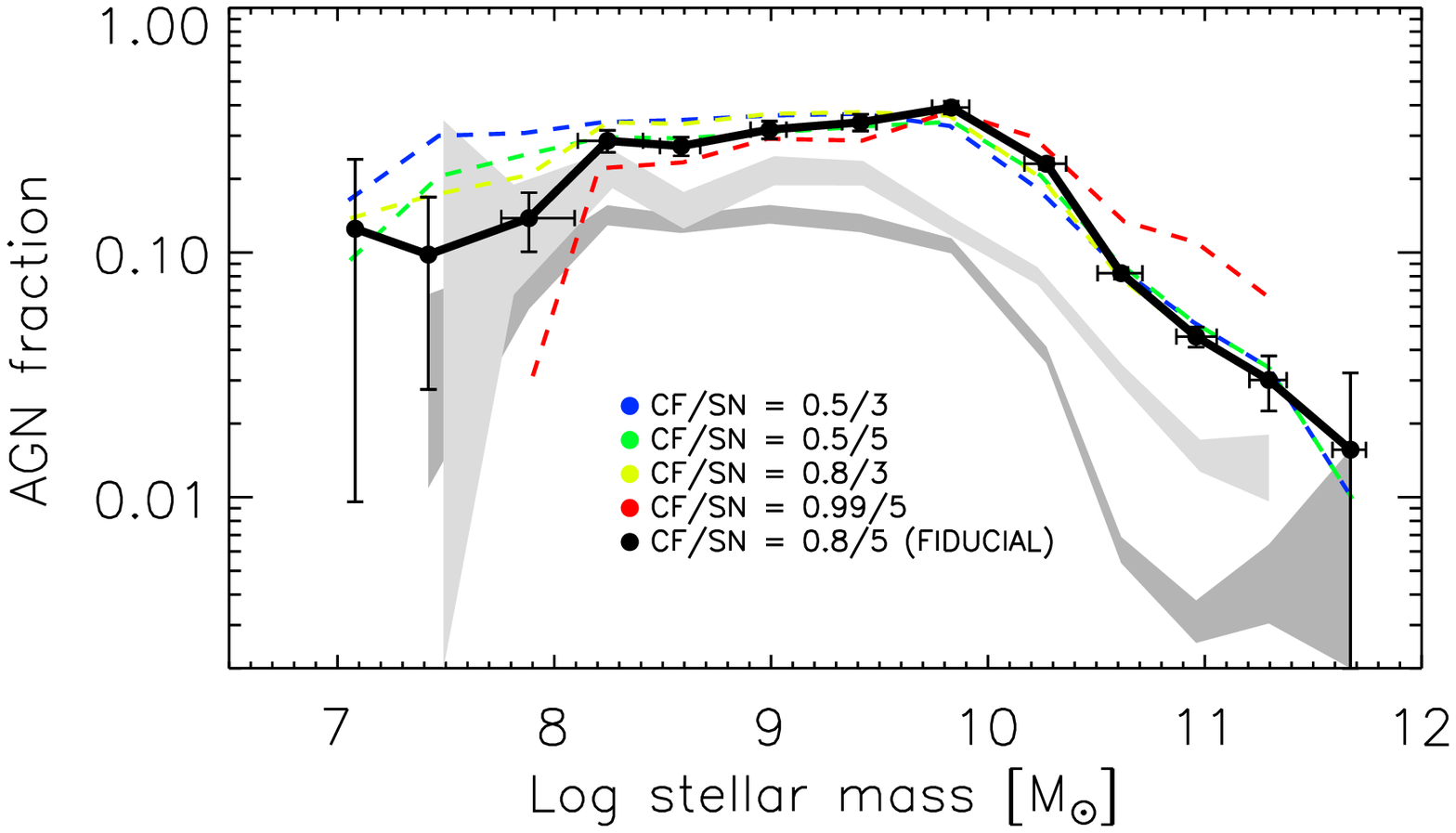}
\caption{\textbf{Left:} $W1-W2$ and $W2-W3$ colours for our HSC galaxy sample. The 2D histogram indicates all galaxies, while the coloured points show the dwarf population, with the colours indicating galaxy stellar masses. The galaxies shown have an S/N greater than 5 in $W1$ and $W2$ and an S/N greater than 2 in $W3$. Common AGN selection methods from \citet[][St12]{Stern2012}, \citet[][J11]{Jarrett2011}, \citet[][S18]{Satyapal2018} and \citet[][S14, which is effectively the criterion used in this study]{Satyapal2014} are shown overplotted. \textbf{Right:} AGN fractions for galaxies with $10^7$ M$_{\odot}$$<$M$_{*}$$<$10$^{12}$ M$_{\odot}$. The solid black line indicates AGN fractions using our fiducial selection criteria, while the dotted coloured lines indicate more relaxed criteria (CF = HSC photo-$z$ confidence parameter, SN = S/N threshold for $W1$ and $W2$ filters). The light and dark grey regions show AGN fractions if the \citet{Stern2012} and \citep{Jarrett2011} classification criteria were used instead. The error bars combine both statistical uncertainties and those due to the WISE photometric errors (estimated via a Monte-Carlo recalculation of the AGN fractions, given the errors in the $W1-W2$ colours).}
\label{fig:AGN fractions}
\end{figure*}

The unprecedented combination of area and depth of the HSC-SSP Wide survey, which is $\sim$4 mags deeper than the SDSS, is useful for producing a statistically significant sample of (the relatively faint) dwarf galaxies. In addition, we employ photometric redshifts, stellar masses and star formation rates (SFRs) in the HSC-SSP DR1 release \citep{Tanaka2018}, that have been derived using the \textsc{mizuki} template-fitting code \citep{Tanaka2015} applied to the optical HSC photometry. 

As described below, we select AGN using infrared data from WISE \citep{Wright2010}, which has mapped the sky at 3.4 ($W1$), 4.6 ($W2$), 12 ($W3$), and 22 ($W4$) microns, with angular resolutions of 6.1, 6.4, 6.5 and 12.0 arcseconds, and 5$\sigma$ point-source sensitivities better than 0.08, 0.11, 1 and 6 mJy respectively. Galaxies from the AllWISE catalog\footnote{http://wise2.ipac.caltech.edu/docs/release/allwise/} are matched to HSC-SSP data, using a search radius of 4 arcseconds.


\subsection{Galaxy sample and infrared selection of AGN}

We select galaxies in the redshift range $0.1<z<0.3$, for which the HSC redshift confidence parameter ($z_{\textnormal{conf}}$) is greater than 0.8 (Figure \ref{fig:completeness}, left panel) and the signal-to-noise (S/N) in the $W1$ and $W2$ filters is greater than 5. 
The lower redshift limit is driven by the fact that the HSC redshifts are most accurate at $z>0.1$ \citep{Tanaka2018}. The upper limit enables us to effectively identify merger-induced tidal features, in order to explore whether AGN in dwarfs are triggered by merging or secular processes. Requiring $z_{\textnormal{conf}}>0.8$ at $z>0.1$ yields median fractional redshift errors in dwarfs of $\sim$9 per cent (inset). The photometric redshifts of dwarfs in this study are of similar quality to those that are routinely employed in massive-galaxy studies at intermediate and high redshifts. It is worth noting that statistical analyses of dwarfs from current and future deep-wide surveys (e.g. LSST) will largely rely on such photometric measurements, since wide-area spectroscopic data of faint galaxies at cosmological distances will be prohibitively expensive, even for the next generation of spectrographs.

Figure \ref{fig:completeness} (right panel) explores the detectability of galaxies in the HSC-SSP as a function of their stellar mass. We plot absolute $r$-band magnitude vs stellar mass, with the grey horizontal lines indicating completeness limits at various redshifts for the HSC-SSP Wide and SDSS spectroscopic datasets, which have $r$-band detection limits of 26 and 22 mag respectively. The HSC-SSP dwarfs agree well with the median locus of local dwarfs from \citet{Woo2008}. For M$_{*} \sim $10$^6$ M$_{\odot}$, no object will typically be visible, even at the lower end of our redshift range ($z \sim 0.1$), while for M$_{*} \sim $10$^9$ M$_{\odot}$, all galaxies should be detected in HSC-SSP Wide imaging out to $z \sim 0.3$. In comparison, the SDSS spectroscopic sample is much shallower, so that even at $z\sim0.1$, typical dwarfs will not be detectable (e.g. an M$_{*}$ $\sim$ 10$^8$ M$_{\odot}$ galaxy that lies on the \citet{Woo2008} locus will appear in the SDSS spectroscopic sample only at $z \sim 0.01$). Note that, since the AGN fraction in galaxies that are undetected should be similar to that in their detected counterparts, incompleteness should not alter the conclusions of this study.

We select AGN using WISE infrared photometry. The hard radiation field around both obscured and unobscured AGN creates hot dust that produces infrared colours that are distinguishable from both stars and galaxies \citep{Jarrett2011}. Recent studies have widely employed AGN classification schemes based on infrared colours from empirical AGN templates (see e.g. left panel in Figure \ref{fig:AGN fractions}). For example, a criterion based on the $W1-W2$ colour \citep[e.g.][]{Stern2012,Satyapal2014}, possibly refined using a further constraint on $W2-W3$ \citep[e.g.][]{Jarrett2011}, is able to identify AGN with high reliability, although adopting a very red colour (e.g. $W1-W2>0.8)$ is likely to miss a significant fraction of bona-fide lower-luminosity AGN \citep[][S18 hereafter]{Satyapal2018}. 

S18 have recently conducted a comprehensive theoretical study of the impact of AGN and extreme starbursts on infrared colours. Their goal was to understand the impact of factors such as the shape of the ionizing radiation field and the properties of the gas and dust (geometry, density, equation of state, chemical composition and grain sizes) on infrared colours, in order to quantify the contamination of AGN selection by starbursts. They show that, at \textit{low} redshift, even extreme starbursts can only mimic AGN-like colours for unrealistically high ionization parameters or gas densities that are inconsistent with the properties of nearby galaxies (the situation may be different in the conditions that are more prevalent at high redshift \citep{Hainline2016}). This yields a wedge in $W1-W2$ vs. $W2-W3$ space that is more inclusive than previous empirical criteria, but selects AGN with high confidence, excluding even vigorously star-forming galaxies at low redshift. 

We note, however, that, given the shape of the S18 wedge, a simpler criterion ($W1-W2$ $>$ 0.52) yields a similar route to estimating AGN fractions, because a negligible number of galaxies have $W2-W3$ colours redward of the wedge at $W1-W2$ $>$ 0.52 (left panel of Figure \ref{fig:AGN fractions}). The benefit of using only $W1$ and $W2$ to select AGN is that many more galaxies have high S/N photometry in these filters compared to $W3$, putting the results on a firmer statistical footing. Thus, we use $W1-W2$ $>$ 0.52 to select AGN in this study (831 of which are in dwarfs), noting that similar criteria have already been used in previous studies \citep{Satyapal2014}. Finally, we note that our AGN show no systematic offset, either in their derived rest-frame photometry (Figure \ref{fig:completeness}) or in their SFRs (not shown), from non-AGN, indicating that AGN do not contaminate the optical SFR measurements. 

\begin{figure}
\begin{center}
\includegraphics[width=0.9\columnwidth]{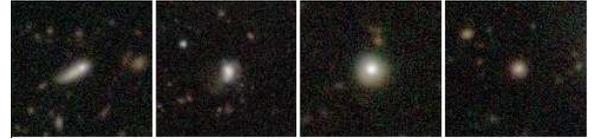}
\caption{HSC $gri$ colour images of dwarf galaxies across the redshift range spanned by our sample ($0.1<z<0.3$). The first two images, from the left, show systems that exhibit tidal features indicative of recent mergers, while the other images show galaxies that do not exhibit tidal features.}
\label{fig:images}
\end{center}
\end{figure}

\begin{figure}
\center
\includegraphics[width=0.95\columnwidth]{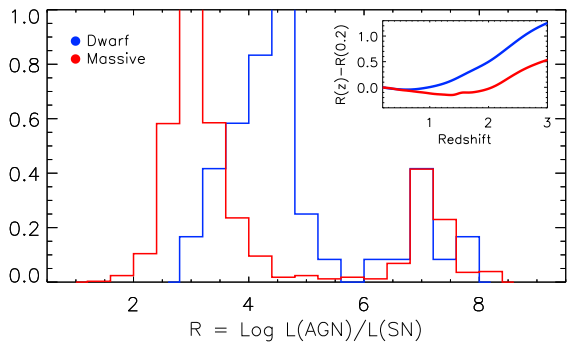}
\caption{\textbf{Main panel:} Log of the ratio of the bolometric AGN luminosity to the total SN luminosity. \textbf{Inset:} The expected evolution of this quantity with redshift, in the Horizon-AGN cosmological simulation for dwarfs and massive galaxies, from \citet{Martin2019}. This ratio is expected to increase with redshift - e.g. at $z\sim3$, the ratio in dwarfs is around a factor of 20 larger compared to its value at $z \sim 0.2$.}
\label{fig:lum ratios}
\end{figure}


\section{AGN fractions and triggering processes}
\label{sec:agn properties}

The right-hand panel of Figure \ref{fig:AGN fractions} shows our AGN fractions as a function of stellar mass. Error bars in the AGN fractions include both the statistical error and that induced by the uncertainties in the $W1$ and $W2$ photometry. In galaxies with stellar masses between 10$^7$ and 10$^9$ M$_{\odot}$, AGN fractions vary between $\sim$10 and $\sim$30 per cent. Not unexpectedly, these AGN fractions are larger than in many previous studies because, as described above, the AGN selection criteria are more inclusive. Relaxing the galaxy selection criteria, e.g. in the HSC redshift confidence parameter or the S/N in the WISE filters, does not significantly alter the overall result (coloured dotted lines). 

The dark and light grey regions show AGN fractions if the \citet{Jarrett2011} and \citet{Stern2012} classification criteria were used instead. {\color{black}Notwithstanding the heterogeneity in techniques that have been used for selecting dwarf AGN, these regions are consistent with AGN fractions reported by previous studies that have used the same criteria across our stellar mass range of interest  \citep[e.g.][]{Satyapal2014,Marleau2017,Goulding2018}. They also exhibit the decrease in infrared-selected AGN fractions between the dwarf and massive-galaxy regimes observed in previous work \citep[e.g.][]{Satyapal2014}.} A significant minority of dwarfs, therefore, appear to host AGN, with AGN fractions that are similar to those in the massive-galaxy regime. For any reasonable duty cycle, this suggests a significant BH-occupation fraction in the dwarf regime. {\color{black}Interestingly, this appears consistent with the dwarf BH-occupation fractions predicted by current simulations \citep[40-70 per cent;][]{Bellovary2019}. 

It is worth noting that, while several studies (including this one), indicate higher infrared-selected AGN fractions in dwarfs than in massive galaxies, the opposite trend tends to be true when AGN are identified via optical or radio wavelengths at low redshift \citep[e.g.][]{Kauffmann2003,Best2005}. This could indicate that the buried phase is shorter in more massive galaxies (which have more powerful AGN), which then manifests itself as a lower AGN fraction in the WISE selection. In addition, selection effects could also play a role. For example, as Figure \ref{fig:completeness} indicates, dwarfs that appear in the SDSS spectroscopic sample at cosmological distances must be anomalously bright, most probably due to high SFRs. These systems will likely be classified as `star-forming' in optical diagnostics (even if they have a buried AGN), producing lower AGN fractions.} 

We proceed by exploring whether the trigger for the AGN activity in dwarfs is likely to be mergers/interactions or secular processes. The deep, high-resolution HSC-SSP Wide imaging (four mags deeper than the SDSS, with around a factor of three better resolution) offers an effective route to identifying systems that have undergone recent mergers, including minor mergers \citep[see e.g.][]{Kaviraj2014a,Kaviraj2014b}. We visually inspect $\sim$400 individual galaxy images of dwarf AGN and non-AGN (see Figure \ref{fig:images}), which have the same redshift and stellar mass distributions, and identify `disturbed' systems, that are either in ongoing mergers or show tidal features indicative of a recent interaction. The disturbed fractions in the dwarf AGN/non-AGN are $6.4^{\pm1.3}$/$7.6^{\pm1.3}$ per cent. Thus, dwarf AGN do not show an excess of disturbed morphologies, indicating that mergers are not a significant driver of AGN activity. The AGN-triggering processes in the dwarf regime are therefore likely to be largely secular in nature. A similar exercise for intermediate ($10^{9}$ M$_{\odot}$$<$M$_{*}$$<$10$^{10.5}$ M$_{\odot}$) and high (M$_{*}$>10$^{10.5}$ M$_{\odot}$) mass galaxies show disturbed fractions for AGN/non-AGN of $10.4^{\pm1.8}$/$11.7^{\pm1.8}$ and $19.1^{\pm2.3}$/$11.6^{\pm1.6}$ per cent respectively. The role of mergers in triggering AGN appears to become progressively more important at higher stellar masses, consistent with other studies in the recent literature \citep[e.g.][]{Shabala2017}. 


\section{AGN vs SN energetics: the plausibility of AGN feedback}
\label{sec:ratios}

We complete our study by comparing the energetics of AGN to that of SN. We estimate bolometric AGN luminosities ($L_{AGN}$) using k-corrected $W3$ fluxes, using the bolometric correction $L_{BOL}$ $\sim$ 12 $\times$ $L_{W3}$ \citep{Richards2006}, which is not strongly dependent on AGN luminosity. K-corrections are calculated using the mean AGN SED template from \citet{Assef2010}. We restrict this analysis to galaxies where the $W3$ S/N is greater than 5. Note that, since the $W3$ S/N is typically much lower than $W1$ or $W2$, these represent the \textit{very brightest} AGN in our sample. The total SN luminosity ($L_{SN}$) can be estimated as $L_{SN}$ = $E_{SN}$.$\phi$.$\nu$, where $E_{SN}$ is the energy output of a Type II SN (10$^{51}$ ergs), $\phi$ is the measured star formation rate and $\nu$ ($\sim$ 1/150 M$_{\odot}$) is the number of SN per unit stellar mass formed \citep[e.g.][]{Dashyan2018}. We note that, in reality, only a fraction of the AGN and SN luminosities are likely to couple to the gas and contribute to feedback. While this coupling efficiency is likely to be 1 to 10 per cent for SN \citep{Dubois2008}, it remains largely unquantified for AGN in dwarfs (although recent studies indicate that the presence of an AGN has a positive impact on the efficiency of feedback in dwarfs, \citep[see e.g.][]{Koudmani2019,Barai2018}). We ignore coupling efficiencies in our analysis here. 

The main panel of Figure \ref{fig:lum ratios} presents $L_{AGN}/L_{SN}$ for both dwarfs and massive galaxies. {\color{black}The observed bimodality is due to the bimodality in the SFR distribution i.e. the existence of a red sequence and a blue cloud. Bolometric luminosities in the dwarf AGN have values of log $L_{AGN}$=46 $\pm$ 0.4 erg s$^{-1}$, consistent with the upper end of values found in observational work \citep[e.g.][]{Woo2002,Mezcua2019}. A canonical 10 per cent efficiency implies accretion rates between 0.03 and 0.2 M$_{\odot}$ yr$^{-1}$. Recall that since we require a $W3$ detection for this analysis, these values likely correspond to the upper end of the $L_{AGN}$ distribution in dwarfs (and appear reasonably consistent with the upper end of predicted $L_{AGN}$ values in current simulations \citep[e.g.][]{McAlpine2018}.}

In all cases, the bolometric AGN luminosities are orders of magnitudes greater than the SN luminosities. Thus, if only a small fraction (less than a percent) of the AGN luminosity is able to couple to the gas reservoir, then it could deliver at least the same level of feedback-driven regulation of star formation as that due to SN. It is worth noting that this surfeit of AGN energy is greater in dwarfs than in their massive counterparts. The potentially large BH occupation fractions in dwarfs, coupled with their high $L_{AGN}/L_{SN}$ ratios, suggests that, if AGN feedback is considered important in the massive galaxy regime, the same may be true in dwarf galaxies, as has been suggested by recent studies \citep[e.g.][]{Penny2018}. The inset shows how this ratio may vary over time, by plotting the ratio between $L_{AGN}/L_{SN}$ at a given redshift and its value at $z\sim0.2$, taken from the Horizon-AGN simulation \citep{Martin2019}. Even though SFRs increase in galaxies towards higher redshifts, resulting in increasing $L_{SN}$, $L_{AGN}$ is expected to increase faster (particularly in dwarfs). Thus, AGN feedback remains plausible throughout cosmic time in the dwarf regime.   


\section{Summary}
\label{sec:summary}

We have combined data from the HSC-SSP and the AllWISE catalog to explore (a) the AGN frequency in dwarf galaxies, (b) the processes that trigger AGN in these systems and (c) the plausibility of AGN feedback in the dwarf regime. Our main conclusions are as follows: 

\begin{itemize}

\item A significant minority of dwarfs host AGN. The AGN fractions in dwarfs with $10^{7}$ M$_{\odot}$$<$M$_{*}$$<$10$^{9}$ M$_{\odot}$ vary between 10 and 30 per cent. Given that the AGN fraction is a lower limit to the BH-occupation fraction, this suggests that, for reasonable duty cycles, a significant fraction of dwarfs are likely to host BHs. 

\item AGN hosts in dwarfs show low merger fractions ($\sim$6 per cent), which are consistent with the merger fractions in non-AGN. This suggests that AGN-triggering processes in dwarfs are largely unrelated to mergers and are predominantly secular in nature. 

\item Bolometric AGN luminosities ($L_{AGN}$) in all galaxies are significantly larger than SN luminosities ($L_{SN}$), with $L_{AGN}/L_{SN}$ being higher in dwarfs than in their massive counterparts. Thus, even if a small fraction of the AGN luminosity can couple to the gas reservoir, this could provide at least as much feedback-driven regulation of star formation as is likely via SN. Furthermore, simulations suggest that $L_{AGN}/L_{SN}$ is expected to increase with redshift. Given the potentially high dwarf BH-occupation fractions, it appears reasonable to suggest that, if AGN feedback is considered important for massive-galaxy evolution, the same is likely to be true in the dwarf regime. 

\end{itemize}

The role of AGN in dwarfs is a key missing puzzle in our understanding of galaxy evolution. Our study demonstrates that the dwarf regime can be probed statistically (using high-quality photometric redshifts alone) via deep-wide surveys like HSC and successor datasets from instruments like LSST. Together with the recent literature, our results indicate that, contrary to our classical thinking, AGN are likely to be important drivers of dwarf-galaxy evolution. 


\section*{Acknowledgements}

We are grateful to the anonymous referee for several constructive suggestions that improved this manuscript. We thank Julien Devriendt and Martin Hardcastle for many interesting discussions. SK acknowledges a Senior Research Fellowship from Worcester College Oxford. GM acknowledges support from the STFC [ST/N504105/1] and a Balzan Fellowship from New College, Oxford. 

The HSC collaboration includes the astronomical communities of Japan, Taiwan and Princeton University. The HSC instrumentation and software were developed by the National Astronomical Observatory of Japan (NAOJ), the Kavli Institute for the Physics and Mathematics of the Universe (Kavli IPMU), the University of Tokyo, the High Energy Accelerator Research Organization (KEK), the Academia Sinica Institute for Astronomy and Astrophysics in Taiwan (ASIAA) and Princeton University. Funding was contributed by the FIRST program from the Japanese Cabinet Office, the Ministry of Education, Culture, Sports, Science and Technology (MEXT), the Japan Society for the Promotion of Science (JSPS), Japan Science and Technology Agency (JST), the Toray Science Foundation, NAOJ, Kavli IPMU, KEK, ASIAA and Princeton University. This paper makes use of software developed for the Large Synoptic Survey Telescope, which is available as free software at  http://dm.lsst.org. It also makes use of data products from the Wide-field Infrared Survey Explorer, which is a joint project of the University of California, Los Angeles, and the Jet Propulsion Laboratory/California Institute of Technology, funded by the National Aeronautics and Space Administration. 

The Pan-STARRS1 Surveys (PS1) have been made possible through contributions of the Institute for Astronomy, the University of Hawaii, the Pan-STARRS Project Office, the Max-Planck Society and its participating institutes, the Max Planck Institute for Astronomy, Heidelberg and the Max Planck Institute for Extraterrestrial Physics, Garching, The Johns Hopkins University, Durham University, the University of Edinburgh, Queen’s University Belfast, the Harvard-Smithsonian Center for Astrophysics, the Las Cumbres Observatory Global Telescope Network Incorporated, the National Central University of Taiwan, the Space Telescope Science Institute, the National Aeronautics and Space Administration under Grant No. NNX08AR22G issued through the Planetary Science Division of the NASA Science Mission Directorate, the National Science Foundation under Grant No. AST-1238877, the University of Maryland, and Eotvos Lorand University (ELTE) and the Los Alamos National Laboratory. Based in part on data collected at the Subaru Telescope and retrieved from the HSC data archive system, which is operated by Subaru Telescope and Astronomy Data Center at National Astronomical Observatory of Japan.


\bibliographystyle{mnras}
\bibliography{references}


\bsp
\label{lastpage}
\end{document}